\documentclass[preprint,preprintnumbers,nobibnotes,pra,showkeys,onecolumn,secnumarabic,showpacs]{revtex4}
\usepackage{amsmath}
\usepackage{graphicx}
\usepackage{dcolumn}
\usepackage{bm}

\input{tcilatex}
\begin{document}

\title{Dynamics of entanglement in the one-dimensional anisotropic XXZ model}
\author{Han Zhang$^{a}$, Yu-Liang Xu$^{b}$, Rong-Tao Zhang$^{a}$, Zhe Wang$%
^{a}$, Pan-Pan Fang$^{a}$, Zhong-Qiang Liu$^{a}$, and Xiang-Mu Kong$^{a,b}$}
\altaffiliation{Corresponding author.\\
E-mail address: kongxm668@163.com(X.-M. Kong).}
\affiliation{$^{a}$College of Physics and Engineering, Qufu Normal University, Qufu
273165, China\\
$^{b}$School of Physics and Optoelectronic Engineering, Ludong University,
Yantai 264025, China}

\begin{abstract}
The dynamics of entanglement in the one-dimensional spin-1/2 anisotropic XXZ
model is studied using the quantum renormalization-group method. We obtain
the analytical expression of the concurrence, for two different quenching
methods, it is found that initial state plays a key role in the evolution of
system entanglement, i.e., the system returns completely to the initial
state every other period$(T=\frac{4\pi }{\sqrt{8+\gamma ^{2}}})$. Our
computations and analysis indicate that the first derivative of the
characteristic time at which the concurrence reaches its maximum or minimum
with respect to the anisotropic parameter occurs nonanalytic behaviors at
the quantum critical point. Interestingly, the minimum value of the first
derivative of the characteristic time versus the size of the system exhibits
the scaling behavior which is the same as the scaling behavior of the system
ground-state entanglement in equilibrium. In particular, the scaling
behavior near the critical point is independent of the initial state.
\end{abstract}

\maketitle

\section{INTRODUCTION}

A fundamental difference between quantum and classical physics is that there
is nonclassical correlation in quantum systems called quantum entanglement 
\cite{1} which not counterpart in classical systems. Therefore, the quantum
entanglement is considered as an important rescurce in quantum information
and quantum computation \cite{1,2}. In addition, quantum entanglement also
plays a significant role in the quantum phase transition (QPT). Due to the
correlation length diverges at the quantum critical points, the entanglement
is a good index of QPTs \cite{3,4,5}. In the past few years, the behavior of
entanglement at the vicinity of quantum critical point has been studied in
various spin models \cite{6,7,8,9,10,11,12,13}, which has achieved great
success in theoretically.

Recent advances in experimental technologies, for instance the cold atoms,
trapped ions \cite{14} and ultrafast pulsed lasers \cite{15}, have made it
possible to study the dynamics of nonequilibrium quantum many-body systems.
In addition, not only studies the entanglement dynamics in many-body spin
systems help us understanding the essence of dynamical quantum phase
transition, but it also provides a possible theoretical reference for the
design of solid-state quantum computer. Therefore, the entanglement dynamics
of many-body quantum systems has attracted widespread attention in recent
years. The entanglement dynamics have been studied in recent years mainly
from the following two different perspectives. On the one hand, some work
studied the propagation of entanglement start from a initial state that the
entanglement has been created in a given portion of the multi-body system 
\cite{16,17,18}. On the other hand, the Hamiltonian changed in time from a
certain ground state of a Hamiltonian $H_{0}$. Recent advances have
demonstrated the effectiveness of a quantum quenching approach for the study
of the entanglement dynamics in many-body spin systems \cite{19,20,21,22,23}%
. For example, the adaptive time-dependent density-matrix renormalization
group was applied to discuss the quantum entanglement dynamics of an open
anisotropic spin-$\frac{1}{2}$ Heisenberg chain \cite{19}. The properties of
entanglement dynamics in the ITM were also studied by the quantum
renormalization group \cite{20}. However, the ground state fidelity and
quench dynamics of the 1D extended quantum compass model in a transverse
field were investigated \cite{22}, which indicate that the fidelity
susceptibility and LE could detect the QPTs in the inhomogeneous system. In
particular, the dynamical quantum phase transitions of an interacting
many-body system has been observed experimentally \cite{21}. Previous
research dedicated to understanding the general properties of nonequilibrium
quantum states and expanding significant concepts such as universality to
nonequilibrium regime. In this paper, we show that the low-energy-state
dynamical quantities of one-dimensional XXZ model can detect the QPTs of the
system, which is universal.

In equilibrium, the entanglement and quantum phase transition of a spin-$%
\frac{1}{2}$ anisotropic Heisenberg chain has been discussed in Ref. \cite{7}%
. In this paper, the entanglement dynamics of one-dimensional XXZ model is
studied using the quantum renormalization group method. It is found that the
entanglement presents cosine variations with time for anisotropic
interaction quench, whereas entanglement is sine variable with time for spin
direction quench. Yet, these two different quench methods correspond to the
same period, which are discussed in detail in the Sec. \ref{A ZA}. In
addition, the evolution behavior of entanglement with respect to anisotropy
parameter, for two quenching methods, is dramatical different, but both of
them appear sigular behavior at the quantum critical point. To gain further
insight, the nonanalytic behavior and scaling behavior of the entanglement
dynamics are studied.

The organization of this paper is as follows. In Sec. \ref{come on} we
introduce the idea of quantum renormalazation group and apply it to the
one-dimensional XXZ model. In Sec. \ref{A ZA} the entanglement dynamics of
spin chain are studied when perform two kinds of typical quench. We
summarize in Sec. \ref{jiayou}.

\section{QUANTUM RENORMALIZATION GROUP\label{come on}}

We introduce the QRG method whose main idea is the elimination or thinning
of the degrees of freedom of the system followed by an iteration. The
purpose of iteration is to gradually reduce the number of variables until a
fixed point is reached. In this paper, the Kadanoff's block approach is
applied to the XXZ model, where the lattice is divided into three sites as a
block. Each block can construct the projection operator onto their lower
eigenvectors. The full Hamiltonian is projected onto these eigenvectors to
obtain the effective Hamiltonian which has structural similarities to the
original Hamiltonian \cite{23,24}. The specific operation is as follows, the
Hamiltonian of the XXZ model on a periodic chain with N sites can be written
as%
\begin{equation}
H(J,\Gamma )=\frac{1}{4}\sum_{i}^{N}\left[ J\left( \sigma _{i}^{x}\sigma
_{i+1}^{x}+\sigma _{i}^{y}\sigma _{i+1}^{y}\right) +\Gamma \sigma
_{i}^{z}\sigma _{i+1}^{z}\right] ,  \label{1a}
\end{equation}%
where $J$ is the exchange coupling constant, $J>0$ corresponds to the
antiferromagnetic system and $J<0$ corresponds to the ferromagnetic system,
here we only study the situation of antiferromagnetic system. $\gamma =\frac{%
\Gamma }{J}$ is the anisotropy parameter, and $\sigma _{i}^{\alpha }$ $%
\left( \alpha =x,y,z\right) $ are Pauli matrices of the $i$th site.

Eq. $\left( \ref{1a}\right) $ can be written as $H=H^{B}+H^{BB}$ by applying
the Kadanoff\textquotedblright s block approach. Here $H^{B}=\sum_{I=1}^{%
\frac{N}{3}}h_{I}^{B}$, with $h_{I}^{B}=\frac{1}{4}[J(\sigma
_{I,1}^{x}\sigma _{I,2}^{x}+\sigma _{I,2}^{x}\sigma _{I,3}^{x}+\sigma
_{I,1}^{y}\sigma _{I,2}^{y}+\sigma _{I,2}^{y}\sigma _{I,3}^{y})+\Gamma
(\sigma _{I,1}^{z}\sigma _{I,2}^{z}+\sigma _{I,2}^{z}\sigma _{I,3}^{z})]$,
is the block Hamiltonian. The $H^{BB}=\frac{1}{4}\sum_{I=1}^{\frac{N}{3}%
}[J(\sigma _{I,3}^{x}\sigma _{I+1,1}^{x}+\sigma _{I,3}^{y}\sigma
_{I+1,1}^{y})+\Gamma \sigma _{I,3}^{z}\sigma _{I+1,1}^{z}]$ is the
interblock Hamiltonian. In terms of matrix product space, the Hamiltonian of
each block $(h_{I}^{B})$ can be exactly diagonalized and get two degenerate
ground states which are used to construct the projection operator $%
(T=\left\vert \varphi _{0}\rangle \langle \Uparrow \right\vert +\left\vert
\varphi _{0}^{\prime }\rangle \langle \Downarrow \right\vert )$. Where $\mid
\varphi _{0}\rangle $ and $\mid \varphi _{0}^{\prime }\rangle $ are the two
degenerate ground states of the block Hamiltonian $(h_{I}^{B})$, $\mid
\Uparrow \rangle $ and $\mid \Downarrow \rangle $ are the renormalization
change effective basis vector of the each block spin operator. Therefore,
the effective Hamiltonian $[H^{eff}=T^{+}(H^{B}+H^{BB})T]$ by the original
Hamiltonian and the projection operator can be written as 
\begin{equation}
H^{eff}=\frac{1}{4}(\sum_{I}^{\frac{N}{3}}[J^{\prime }(\sigma _{I}^{x}\sigma
_{I+1}^{x}+\sigma _{I}^{y}\sigma _{I+1}^{y})+\Gamma ^{\prime }\sigma
_{I}^{z}\sigma _{I+1}^{z})],  \label{2a}
\end{equation}%
where 
\begin{equation}
J^{\prime }=\frac{16J^{3}k^{2}}{(8J^{2}+k^{2})^{2}},\text{ \ }\Gamma
^{\prime }=\frac{k^{4}\Gamma }{(8J^{2}+k^{2})^{2}},\text{ \ }k=\Gamma +\sqrt{%
8J^{2}+\Gamma ^{2}}.  \label{3a}
\end{equation}%
Eq. $\left( \ref{3a}\right) $ is the QRG equations. We define a
dimensionless anisotropy parameter $\gamma =\Gamma /J$ which determines the
phase transition properties of the system. The QRG equations can also be
written as

\begin{equation}
\text{ \ }\gamma ^{\prime }=\frac{\gamma }{16}(\gamma +q)^{2},\text{ \ }q=%
\sqrt{8+\gamma ^{2}}.  \label{4}
\end{equation}
\ \ \ \ \ \ \ \ \ \ \ \ \ \ \ \ \ \ \ \ \ \ \ \ \ \ \ \ \ \ \ \ \ \ \ \ \ \
\ \ \ \ \ \ \ \ \ \ \ \ \ \ \ \ \ 

The stable and unstable fixed points can be obtained by solving $\gamma
\equiv \gamma ^{^{\prime }}\equiv \gamma ^{\ast }$. The stable fixed points
locate at $\gamma =0$ and $\gamma =\infty $, the unstable fixed point is $%
\gamma =1$ which is the critical point of the model. As the number of QRG
iterations increase, starting with any initial values for $\gamma >1$, the
coupling parameter flows toward infinity indicating that the system falls
into the universality class of Ising model. But for $\gamma <1$ the stable
fixed point$(\gamma =0)$ is touched. The model expresses a spin fluid phase
when $0\leq \gamma \leq 1$, $\gamma >1$ the model expresses a N\'{e}el phase.

\section{ENTANGLEMENT DYNAMICS\label{A ZA}}

In this section, the entanglement dynamics of XXZ spin chain is analyzed
when the anisotropic interaction and the spin direction are quenched. The
following these two kinds of quenches process will be discussed in detail.

\subsection{Anisotropic interaction quench}

We consider that the spin chain is initialized on one of the two degenerate
ground states $\mid \varphi _{01}\rangle $ of the XX model. Experimentally,
the initial state can be obtained by adding an otherwise inconsequential
infinitesimal magnetic field to the XXZ model \cite{25}. In terms of matrix
product state \cite{26}, the ground state $\mid \varphi _{01}\rangle (\mid
\varphi _{01}\rangle =\frac{1}{2}\mid \uparrow \uparrow \downarrow \rangle -%
\frac{\sqrt{2}}{2}\mid \uparrow \downarrow \uparrow \rangle +\frac{1}{2}\mid
\downarrow \uparrow \uparrow \rangle )$ of the three-site XX model can be
obtained, here $\mid \uparrow \rangle $ and $\mid \downarrow \rangle $ are
eigenvector of $\sigma ^{z}$. The anisotropic interaction parameter suddenly
increases from zero when time $t=0$. In other words, the Hamiltonian is
suddenly converted from $H_{01}$ into $H$, where $H$ is the Hamiltonian of
the XXZ model. The system state evolves to $\mid \varphi _{1}(t)\rangle
=e^{-iHt}\mid \varphi _{01}\rangle $, here 
\begin{equation}
\mid \varphi _{1}(t)\rangle =a1\mid \uparrow \uparrow \downarrow \rangle
+b1\mid \uparrow \downarrow \uparrow \rangle +a1\mid \downarrow \uparrow
\uparrow \rangle ,  \label{4a}
\end{equation}%
where 
\begin{eqnarray}
a1 &=&\frac{e^{\frac{1}{4}iJ\gamma t}[q^{2}\cos (\frac{1}{4}Jqt)-i(\gamma -2%
\sqrt{2})q\sin (\frac{1}{4}Jqt)]}{2q^{2}},\text{ }  \label{5a} \\
\text{\ }b1 &=&-\frac{e^{\frac{1}{4}iJ\gamma t}[\sqrt{2}q^{2}\cos (\frac{1}{4%
}Jqt)+i(4+\sqrt{2}\gamma )q\sin (\frac{1}{4}Jqt)]}{2q^{2}}.
\end{eqnarray}%
Thus that the pure-state density matrix of the three-site system at time $t$
is defined by

\begin{equation}
\rho _{1}(t)=\mid \varphi _{1}(t)\rangle \langle \varphi _{1}(t)\mid .
\label{6a}
\end{equation}%
In the product space of $\sigma ^{z}$, $\rho _{1}(t)$ can be written as

\begin{equation}
\rho _{1}(t)=\left[ 
\begin{array}{cccccccc}
0 & 0 & 0 & 0 & 0 & 0 & 0 & 0 \\ 
0 & w & x & 0 & w & 0 & 0 & 0 \\ 
0 & x^{\ast } & y & 0 & x^{\ast } & 0 & 0 & 0 \\ 
0 & 0 & 0 & 0 & 0 & 0 & 0 & 0 \\ 
0 & w & x & 0 & w & 0 & 0 & 0 \\ 
0 & 0 & 0 & 0 & 0 & 0 & 0 & 0 \\ 
0 & 0 & 0 & 0 & 0 & 0 & 0 & 0 \\ 
0 & 0 & 0 & 0 & 0 & 0 & 0 & 0%
\end{array}%
\right] ,  \label{7a}
\end{equation}%
the expectation values of Pauli matrices and its correlation functions can
be represented by the time-dependent density matrix,%
\begin{eqnarray}
\left\langle \sigma _{1}^{z}\right\rangle &=&y=\frac{1}{2}+\frac{2\sqrt{2}%
\gamma \sin ^{2}(\frac{1}{4}Jqt)}{q^{2}},  \label{8a} \\
\left\langle \sigma _{1}^{z}\sigma _{3}^{z}\right\rangle &=&y-2w=-\frac{4%
\sqrt{2}\gamma \sin ^{2}(\frac{1}{4}Jqt)}{q^{2}},  \label{8b} \\
\left\langle \sigma _{1}^{x}\sigma _{2}^{x}\right\rangle &=&x+x^{\ast }=-%
\frac{8+\gamma ^{2}\cos (\frac{1}{2}Jqt)}{\sqrt{2}q^{2}},  \label{8c} \\
\left\langle \sigma _{1}^{x}\sigma _{2}^{y}\right\rangle &=&ix-ix^{\ast }=-%
\frac{\gamma \sin (\frac{1}{2}Jqt)}{\sqrt{2}q}.  \label{8d}
\end{eqnarray}%
When $\gamma $ is a fixed value, it can be seen that the mean value of $%
\sigma _{1}^{z}$ and its correlation function are all periodic functions
with respect to time based on Eqs. $\left( \ref{8a}\right) $ to $\left( \ref%
{8d}\right) $. Similarly, when $\gamma $ is a fixed value, each matrix
element of $\rho _{1}(t)$ is also periodic function with respect to time
because that each matrix element can be regarded as a function of the
expectation value of Pauli matrix and its correlation function, as follows 
\begin{eqnarray}
w &=&\frac{1}{2}(\left\langle \sigma _{1}^{z}\right\rangle -\left\langle
\sigma _{1}^{z}\sigma _{3}^{z}\right\rangle )=\frac{1}{4}-\frac{\sqrt{2}%
\gamma \sin ^{2}(\frac{1}{4}Jqt)}{q^{2}},\text{ }  \label{88a} \\
\text{\ }x &=&\frac{1}{2}(\left\langle \sigma _{1}^{x}\sigma
_{2}^{x}\right\rangle -i\left\langle \sigma _{1}^{x}\sigma
_{2}^{y}\right\rangle )=-\frac{8+\gamma ^{2}\cos (\frac{1}{2}Jqt)-i\gamma
q\sin (\frac{1}{2}Jqt)}{2\sqrt{2}q^{2}},\text{ } \\
\text{\ }y &=&-\left\langle \sigma _{1}^{z}\right\rangle =\frac{1}{2}+\frac{2%
\sqrt{2}\gamma \sin ^{2}(\frac{1}{4}Jqt)}{q^{2}}.
\end{eqnarray}

It is well known that there are many measurement methods for pairwise
entanglement \cite{30,31,32,33}. In this paper, we calculate the concurrence
of the system and observe the evolution rules of concurrence with time. In
order to without loss of generality, we trace over site $2$. The reduced
density matrix for sites 1 and 3 can be obtained as

\begin{equation}
\rho _{13}(t)=\left[ 
\begin{array}{cccc}
y & 0 & 0 & 0 \\ 
0 & w & w & 0 \\ 
0 & w & w & 0 \\ 
0 & 0 & 0 & 0%
\end{array}%
\right] .  \label{9a}
\end{equation}%
The concurrence between the sites 1 and 3 is defined as%
\begin{equation}
C_{1}(t)=\max \{\sqrt{\lambda _{4}}-\sqrt{\lambda _{3}}-\sqrt{\lambda _{2}}-%
\sqrt{\lambda _{1}},0\},  \label{10a}
\end{equation}%
where the $\lambda _{k}(k=1,2,3,4)$ are the eigenvalues of $\widehat{R}=\rho
_{13}(t)\widetilde{\rho }_{13}(t)$[with $\widetilde{\rho }_{13}(t)=(\sigma
_{1}^{y}\otimes \sigma _{3}^{y})\rho _{13}^{\ast }(\sigma _{1}^{y}\otimes
\sigma _{3}^{y})$ is the spin-flipped density matrix operator] in ascending
order. The eigenvalues of $\widehat{R}$ can be accurately solved:%
\begin{equation}
\lambda _{1}=\lambda _{2}=\lambda _{3}=0,~~\lambda _{4}=4w^{2}.  \label{11a}
\end{equation}%
Therefore, the concurrence of the three-site system corresponding to the
first quench procedure is obtained 
\begin{equation}
C_{1}(t)=2w=\left\langle \sigma _{1}^{z}\right\rangle -\left\langle \sigma
_{1}^{z}\sigma _{3}^{z}\right\rangle =\frac{1}{2}-\frac{2\sqrt{2}\gamma \sin
^{2}(\frac{1}{4}Jqt)}{q^{2}}.  \label{12a}
\end{equation}%
When $\gamma $ is a fixed value, the $C_{1}(t)$ is a periodic function with
respect to time based on Eq. $\left( \ref{12a}\right) $. The concurrence
between spin blocks whose with different number of sites can be obtained by
step by step iteratively QRG equation.

\subsection{Spin $z$ axis rotation quench}

The three-site spin system is initialized on the ground state of Hamiltonian 
$H_{02}(H_{02}=-\frac{J}{4}[\sigma _{1}^{x}\sigma _{2}^{x}+\sigma
_{2}^{x}\sigma _{3}^{x}+\sigma _{1}^{y}\sigma _{2}^{y}+\sigma _{2}^{y}\sigma
_{3}^{y}-\gamma (\sigma _{1}^{z}\sigma _{2}^{z}+\sigma _{2}^{z}\sigma
_{3}^{z})])$ that it is obtained by rotating a $\pi $ around the $z$ axix
for all even sites and leave all odd sites unchanged in the XXZ model, which
can be achieved by pulsed laser in experimentally \cite{25}. The initial
state $\mid \varphi _{02}\rangle (\mid \varphi _{02}\rangle =\frac{d}{2}\mid
\uparrow \uparrow \downarrow \rangle -\frac{q+\gamma }{2}\mid \uparrow
\downarrow \uparrow \rangle +\frac{d}{2}\mid \downarrow \uparrow \uparrow
\rangle )$ evolves under the Hamiltonian $H$ of XXZ system when time $t=0$,
where $d=\sqrt{8+(q+\gamma )^{2}}$. The state of the system evolves to $\mid
\varphi _{2}(t)\rangle =e^{-iHt}\mid \varphi _{02}\rangle $. Thus that the
time-dependent density matrix of three-site system can be written as%
\begin{equation}
\rho _{2}(t)=\mid \varphi _{2}(t)\rangle \langle \varphi _{2}(t)\mid .
\label{13a}
\end{equation}

As we have mentioned above, each matrix element of $\rho _{2}(t)$ can also
be derived from the average value of the Pauli matrix and its correlation
functions, which is also periodic function with respect to time. We obtain
the concurrence of sites $1$ and $3$ using the previous method. The
concurrence is%
\begin{equation}
C_{2}(t)=\frac{8\gamma -\gamma ^{3}+q^{3}-16\gamma \cos (\frac{1}{2}Jqt)}{%
2q^{3}}.  \label{14a}
\end{equation}%
Obviously, $C_{2}(t)$ is a periodic function with respect to time when $%
\gamma $ is a fixed value. Similarly, entanglement between larger spin
blocks can be obtained by QRG equation. For simplicity and without loss of
generality, we will choose the exchange coupling $J=1.0$ afterwards.

\subsection{Evolution of concurrence}

It is easy to see from Eq. $\left( \ref{12a}\right) $ and $\left( \ref{14a}%
\right) $ that the concurrence of the system mainly depend on the time$(t)$
and the anisotropy parameter$(\gamma )$. For different QRG steps, $C_{1}(t)$
and $C_{2}(t)$ versus time$(t)$ for the different values of $\gamma $ is
plotted in Fig. 1. As can be seen from Figs. 1(a1) and (a2), the $C_{1}(t)$
shows a cosine variations with time while $C_{2}(t)$ shows a sine change
with time, but $C_{1}(t)$ and $C_{2}(t)$ have the same period$(T=\frac{4\pi 
}{J\sqrt{8+\gamma ^{2}}})$ with time. As shown in Fig. 1(a1) and (a2), as
the size of the system increases, the lowest peak of $C_{1}(t)$ gradually
becomes higher when $\gamma =0.9$, conversely, each peak of $C_{2}(t)$
gradually decreases. It is interesting that the periods of $C_{1}(t)$ and $%
C_{2}(t)$ progressively become larger under QRG iteration and finally $%
C_{1}(t)$ and $C_{2}(t)$ are equal to $0.5$ in the thermodynamic limit. As
we have mentioned previously, the coupling constant flows to the stable
fixed point$(\gamma =0)$ under QRG iteration when $\gamma =0.9$. At $\gamma
=0$, the initial Hamiltonian$(H_{01})$ and the evolutionary Hamiltonian$(H)$
are the same so that the $C_{1}(t)$ does not change with time. For $\gamma
=1 $ (see the left insets of Fig. 1(a1) and (a2)), the invariance between
the concurrence of different-length chains is the result of the correlation
length divergence at $\gamma _{c}=1$. For $\gamma =1.1$ (see in Fig. 1(a1)
right inset), when the QRG iteration tends to infinity, the lowest peak of $%
C_{1}(t)$ decreases at first and then increases and finally oscillates
around $0.5$. In Fig. 1(a2) right inset, as the size of the system
increases, the height of each peak of $C_{2}(t)$ increases at first and then
decreases gradually and finally vanishes when $N\rightarrow \infty $. It is
easy to see that increasing the length of the chain shortens the periods of $%
C_{1}(t)$ and $C_{2}(t)$ when $\gamma =1.1$.

In order to further understand the evolution of entanglement, we calculate
that the probability of the evolved ground state returns to the initial
ground state for two different quench types, separately \cite{27,28}. As
follows%
\begin{equation}
P_{1}=\left\vert \langle \varphi _{01}\mid \varphi _{1}(t)\rangle
\right\vert ^{2}=\frac{16+\gamma ^{2}+\gamma ^{2}\cos (\frac{1}{2}qt)}{%
16+2\gamma ^{2}},  \label{15a}
\end{equation}%
\begin{equation}
P_{2}=\left\vert \langle \varphi _{02}\mid \varphi _{2}(t)\rangle
\right\vert ^{2}=\frac{64+\gamma ^{4}+16\gamma ^{2}\cos (\frac{1}{2}qt)}{%
q^{4}}.  \label{16a}
\end{equation}%
Higher values of $P_{1}$ and $P_{2}$ mean that the system is easier to
return to the initial state, in addition, the system completely returns its
initial state when $P_{1}=1$ or $P_{2}=1$. As can be seen from Eq. $\left( %
\ref{15a}\right) $ and $\left( \ref{16a}\right) $, for a fixed value of $%
\gamma $, $P_{1}$ and $P_{2}$ are periodic functions with respct to time and
the periods are $T=\frac{4\pi }{\sqrt{8+\gamma ^{2}}}$. It is means that the
system will completely return to the initial state every other period. In
particular, the greater the $\gamma $, the shorter the period.

For different QRG steps, the evolution of $C_{1}(t)$ and $C_{2}(t)$ versus $%
\gamma $ for $t=11.5$ and $t=1.5$ is plotted in Fig. 2. We find that the
changes of concurrence corresponding to the two quenching methods with
respect to anisotropy parameters are different, it is means that initial
state plays an important role in the evolution of system entanglement.
Moreover, the short-time (sufficiently near to the initial moment) behavior
and long-time (far from the initial moment) behavior of concurrence are
somewhat different. As shown in Fig. 2(b1), at $t=11.5$, the concurrence
decreases from the equilibrium state to a finite value and then start to
oscillate when the $\gamma $ is turned on. However, at $t=1.5$ (see in Fig.
2(b1) inset), the concurrence decays from equilibrium state to zero and then
begin to oscillate. As $\gamma $ increases, the values of each trough of $%
C_{1}(t)$ gradually high and finally severe oscillates around $0.5$ when $%
\gamma $ tend to infinity. It is because that the period of the system
returns completely to the initial state approaches zero when $\gamma $ tends
to infinity, which can be obtained from Eq. $\left( \ref{15a}\right) $,
hence, $C_{1}(t)$ violently oscillates around $0.5$ when $\gamma \rightarrow
\infty $ because that the ground state entanglement of the initial
Hamiltonian$(H_{01})$ equals $0.5$ and it is independent of $\gamma $. The
changes of $C_{2}(t)$ versus $\gamma $ is different from $C_{1}(t)$ when $%
t=11.5$ and $t=1.5$. From the Fig. 2(b2), as $\gamma $ increases, we find
that the $C_{2}(t)$ increases from the initial state to a finite value and
then begins to oscillate, but for $t=1.5$ $\left[ \text{see the inset of
Figs. 2(b2)}\right] $, the $C_{2}(t)$ reaches its maximum at first and then
begins to oscillate. the height of each peak of $C_{2}(t)$ gradually
decreases with $\gamma $ increases and finally vanishes as $\gamma
\rightarrow \infty $, it is because that the ground state entanglement of $%
H_{02}$ is related to $\gamma $, i.e., the ground state entanglement of $%
H_{02}$ tends to zero when $\gamma \rightarrow \infty $. Although the change 
$C_{1}(t)$ and $C_{2}(t)$ versus $\gamma $ are different, but they all
reveal that increasing the length of the chain enhances the oscillation of
the $C_{1}(t)$ and $C_{2}(t)$. Moreover, in the thermodynamic limit, $%
C_{1}(t)$ and $C_{2}(t)$ all happen mutation at the critical point because
that the result of the correlation length divergence at $\gamma _{c}=1$.

The non-analytic behavior of a physical quantity is a characteristic of QPT.
The non-analytic behavior is often accompanied by a scaling behavior since
that the correlation length diverges at the critical point. In this section,
we demonstrate that the characteristic time can be used to describe the
critical phenomenon of the one-dimensional anisotropic XXZ model which in
the vicinity of the transition point. For any the anisotropy parameter, we
define the characteristic time $T_{\min }^{k}(\gamma )$ at which the $%
C_{1}(t)$ reaches its $k$th minimum values and $T_{\max }^{k}(\gamma )$ at
which the $C_{2}(t)$ reaches its $k$th maximum values. The characteristic
time is analyzed as a function of the coupling constant$(\gamma )$ at
different QRG steps.

Further insight, we analyze the first derivatives of $T_{\min }$ and $%
T_{\max }$ with respect to the coupling constant$(\gamma )$ for $k=1$ in
Fig. 3, which show the singular behavior at the critical point as the size
of the system becomes large. The inset of Fig. 3 are the change of $T_{\min
} $ and $T_{\max }$ versus $\gamma $ at different QRG steps, which shows
that $T_{\min }$ and $T_{\max }$ develop two saturated values in the
thermodynamic limit. Specially, as shown in Fig. 3(c1) and Fig. 3(c2), $%
\frac{dT_{\min }}{d\gamma }$ and $\frac{dT_{\max }}{d\gamma }$ vs the $%
\gamma $ are both of the same singular behavior because that the period of $%
C_{1}(t)$ and $C_{2}(t)$ are identical when the $\gamma $ is a constant. For
a more detailed analysis, the position of the minimum ($\gamma _{m}$) of $%
\frac{dT_{\min }}{d\gamma }$ approaches the critical point with the size of
the system increase. This is plotted in Fig. 4, which shows the relation $%
\gamma _{m}=\gamma _{c}+N^{-\theta }$ with $\theta =0.47$. Besides, the
scaling behavior of $y\equiv \left\vert \frac{dT_{\min }}{d\gamma }%
\right\vert _{\gamma _{m}}$ versus $N$ is plotted in Fig. 5 which shows a
liner behavior of $\ln (y)$ versus $\ln (N)$, i.e., $\left\vert \frac{%
dT_{\min }}{d\gamma }\right\vert _{\gamma _{m}}\sim N^{0.46}$. Moreover, the
exponent $\theta $ is directly related to the correlation length exponent $%
\nu $ at the vicinity of critical point $(\gamma _{c})$, i.e., the relation
is $\theta =\frac{1}{\nu }$. Interestingly, the characteristic time
represents scaling behavior close to the quantum critical point with
exponent $\theta =0.47$ which fantastic corresponds to the entanlement
exponent of the one-dimensional XXZ model. Remarkably, the scaling behavior
of $\left\vert \frac{dT_{\max }}{d\gamma }\right\vert _{\gamma _{m}}$ versus 
$N$ is the same as $\left\vert \frac{dT_{\min }}{d\gamma }\right\vert
_{\gamma _{m}}$ because that $C_{1}(t)$ and $C_{2}(t)$ have the same period
when $\gamma $ is a constant. Therefore, we only study one of them.

\section{SUMMARY\label{jiayou}}

In this paper, the dynamics of entanglement for the one-dimensional spin-$%
\frac{1}{2}$ anisotropic XXZ model are studied using the quantum
renormalization-group method. We obtain the analytic expressions of
concurrence of the system corresponding to two different quenching methods.
We find that the initial state plays a key role in the evolution of system
entanglement. In order to further understand the dynamics of system
entanglement, we investigate the probabilities when the system return to the
initial state Corresponding to the two quenching methods, the result shows
that both of them return completely to the initial state with period$(T=%
\frac{4\pi }{\sqrt{8+\gamma ^{2}}})$. The period is related to the
anisotropy parameter, i.e., when $\gamma \rightarrow \infty $, the period
when the system returns completely to the initial state approaches zero. We
demonstrate that the characteristic time can detect the QPT of
one-dimensional XXZ model. Interestingly, we find that the scaling behavior
of $\left\vert \frac{dT_{\max }}{d\gamma }\right\vert _{\gamma _{m}}$ versus 
$N$ close to the critical point are similar to those of the XXZ model in
equilibrium and find the scaling behavior is independent of the initial
state or quenching method.

\begin{acknowledgments}
This work was supported by the National Natural Science foundation of China
under Grant NO. 11847086, NO. 11675090.
\end{acknowledgments}

\newpage

\begin{center}
\textbf{Figure Captions}
\end{center}

Fig.~$1$ (Color online) For different QRG steps, $C_{1}(t)$ and $C_{2}(t)$
change with time $t$. Where, the left inset corresponds to $\gamma =1$ and
the right panel corresponds to $\gamma =1.1$.

\bigskip

Fig.~$2$ (Color online) Evolution of the concurrence versus $\gamma $ for
different values of time in terms of QRG iterations. The inset shows the
change of $C_{1}(t)$ and $C_{2}(t)$ with anisotropy parameters at $t=1.5$
and Figs. 2. (b1) and (b2) correspond to the $t=11.5$, which is relatively
large compared with $1.5$.

\bigskip

Fig.~$3$ (color online) First derivative of $T_{\min }$ and $T_{\max }$ and
their manifestation toward diverging as the number of QRG iteration
increase. Inset: Evolution of $T_{\min }$ and $T_{\max }$ with respect to
the anisotropy parameter as the number of QRG iterations increase.

\bigskip

Fig. $4$ The scaling behavior of $\gamma _{m}$ for different-length chains,
where, $\gamma _{m}$ is the position of minimum of $\frac{dT_{\min }}{%
d\gamma }$ in Fig. 3(c1).

\bigskip

Fig. $5$ The scaling behavior of $\left\vert \frac{dT_{\min }}{d\gamma }%
\right\vert _{\gamma _{m}}$ versus system size $N$.


\begin{thebibliography}{99}
\bibitem{1} J. S. Bell, Physics (Long Island City, NY) \textbf{1}, 195
(1964).

\bibitem{2} M. A. Nielsen and I. L. Chuang, Quantum Computation and Quantum
Communication (Cambridge University Press, Cambridge, 2000).

\bibitem{3} A. Osterloh, L. Amico, G. Falci and R. Fazio, Nature (London) 
\textbf{416, }608 (2002).

\bibitem{4} S. Sachdev, Quantum Phase Transitions (Cambridge University
Press, Cambridge, 2000).

\bibitem{5} T. J. Osborne and M. A. Nielsen, Phys. Rev. A \textbf{66, }%
032110 (2002).

\bibitem{6} M. Kargarian, R. Jafari and A. Langari, Phys. Rev. A \textbf{76, 
}060304 (2007).

\bibitem{7} M. Kargarian, R. Jafari and A. Langari, Phys. Rev. A \textbf{77, 
}032346 (2008).

\bibitem{8} F.-W. Ma, S.-X. Liu and X.-M. Kong, Phys. Rev. A \textbf{83, }%
062309 (2011); \textbf{84, }042302 (2011).

\bibitem{9} Y.-L. Xu, X.-M. Kong, Z.-Q. Liu and C.-Y. Wang, Phys. A
(Amsterdam, Neth.) \textbf{446}, 217 (2016).

\bibitem{10} Y.-L. Xu, L.-S. Wang and X.-M. Kong, Phys. Rev. A \textbf{87, }%
012312 (2013).

\bibitem{11} M. Kargarian, R. Jafari and A. Langari, Phys. Rev. A \textbf{%
79, }042319 (2009).

\bibitem{12} Y.-L. Xu, X. Zhang, Z.-Q. Liu, X.-M. Kong and T.-q. Ren, Eur.
Phys. J. B \textbf{87}, 132 (2014).

\bibitem{13} Y.-L. Xu, X.-M. Kong, Z.-Q. Liu and C.-C Yin, Phys. A \textbf{95%
}, 042327 (2017).

\bibitem{14} A. Lamacraft and J. Moore, Ultracold Bosonic and Fermionic
Gases (Elsevier, Oxford, UK, 2012), Vol. 5.

\bibitem{15} N. Gedik, D.-S. Yang, G. Logvenov, I. Bozovic and A. H. Zewail,
Science \textbf{316, }425 (2007).

\bibitem{16} Amico, L., A. Osterloh, F. Plastina, G. Palma and R. Fazio,
Phys. Rev. A \textbf{69, }022304 (2004).

\bibitem{17} S. D. Hamieh and M. I. Katsnelson, Phys. Rev. A \textbf{72, }%
032316 (2005)\textbf{.}

\bibitem{18} M. J. Hartmann, M. E. Reuter and M. B. Plenio, New J. Phys. 
\textbf{8, }94 (2006).

\bibitem{19} Jie Ren and Shiqun Zhu, Phys. Rev. A \textbf{81, }014302 (2010).

\bibitem{20} R. Jafari, Phys. Rev. A \textbf{82, }052317 (2010).

\bibitem{21} P. Jurcevic, H. Shen, P. Hauke, C. Maier, T. Brydges, C.
Hempel, B. P. Lanyon, M. Heyl, R. Blatt and C. F. Roos, Phys. Rev. L \textbf{%
119, }080501 (2017).

\bibitem{22} R Jafari, J. Phys. A \textbf{49, }185004 (2016).

\bibitem{23} J. Eisert, M. Friesdorf and C. Gogolin, Nat. Phys. \textbf{11, }%
124 (2015).

\bibitem{24} M. A. Martin-Delgado and G. Sierra, Int. J. Mod. Phys. A 
\textbf{11, }3145 (1996).

\bibitem{25} A. Langari, Phys. Rev. B \textbf{58, }14467 (1998).

\bibitem{26} R. A. Kaden, Hazzard, V. D. W. Mauritz, F.-F. Michael, R.
Salvatore, Manmana, G. Emanuele, T. Dalla, P. Tilman, K. Michael and M. R.
Ana, Phys. Rev. A \textbf{90, }063622 (2014)\textbf{.}

\bibitem{27} F. Verstraete, J. I. Cirac, J. I. Latorre, E. Rico and M. M.
Wolf, Phys. Rev. Lett. \textbf{94, }140601\textbf{\ }(2005).

\bibitem{28} A. Rajak and U. Divakaran, J. Stat. Mech. P04023 (2014).

\bibitem{29} A. Dutta, G. Aeppli, B. K. Chakrabarti, U. Divakaran, T.
Rosenbaum and D. Sen, Quantum Phase Transitions in Transverse Field Spin
Models: From Statistical Physics to Quantum Information (Cambridge:
Cambridge University Press, 2015).

\bibitem{30} Y. X. Chen and D. Yang, Quant. Info. Proc.\textbf{\ 1, }389
(2003).

\bibitem{31} G. Vidal and R. F. Werner, Phys. Rev. A \textbf{65, }032314
(2002).

\bibitem{32} E. M. Rains, Phys. Rev. A\textbf{\ 60, }179 (1999).

\bibitem{33} S. Hill and W. K. Wootters, Phys. Rev. Lett. \textbf{78,} 5022
(1997);W. K. Wootters, ibid. \textbf{80, }2245 (1998).
\end{thebibliography}
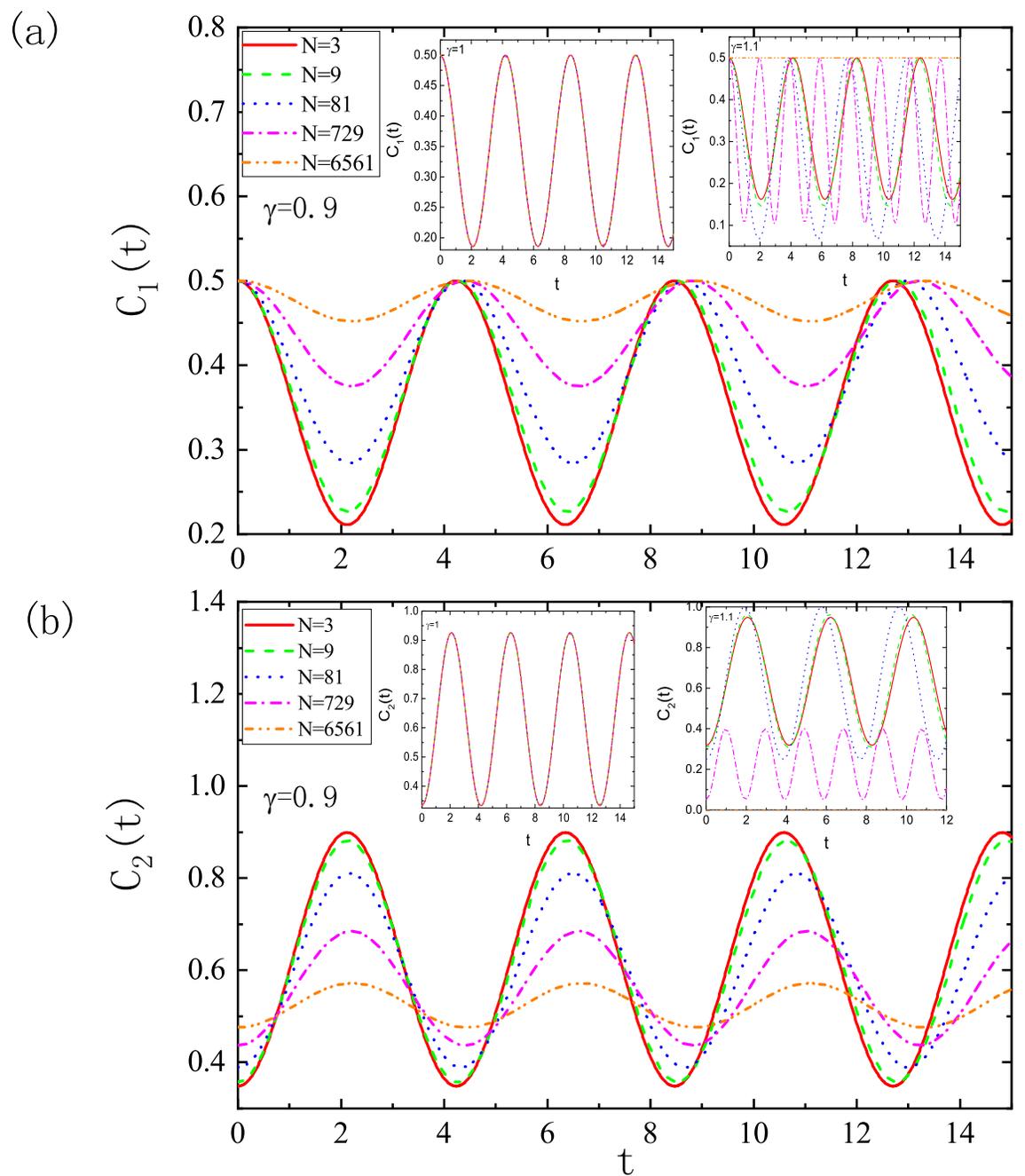
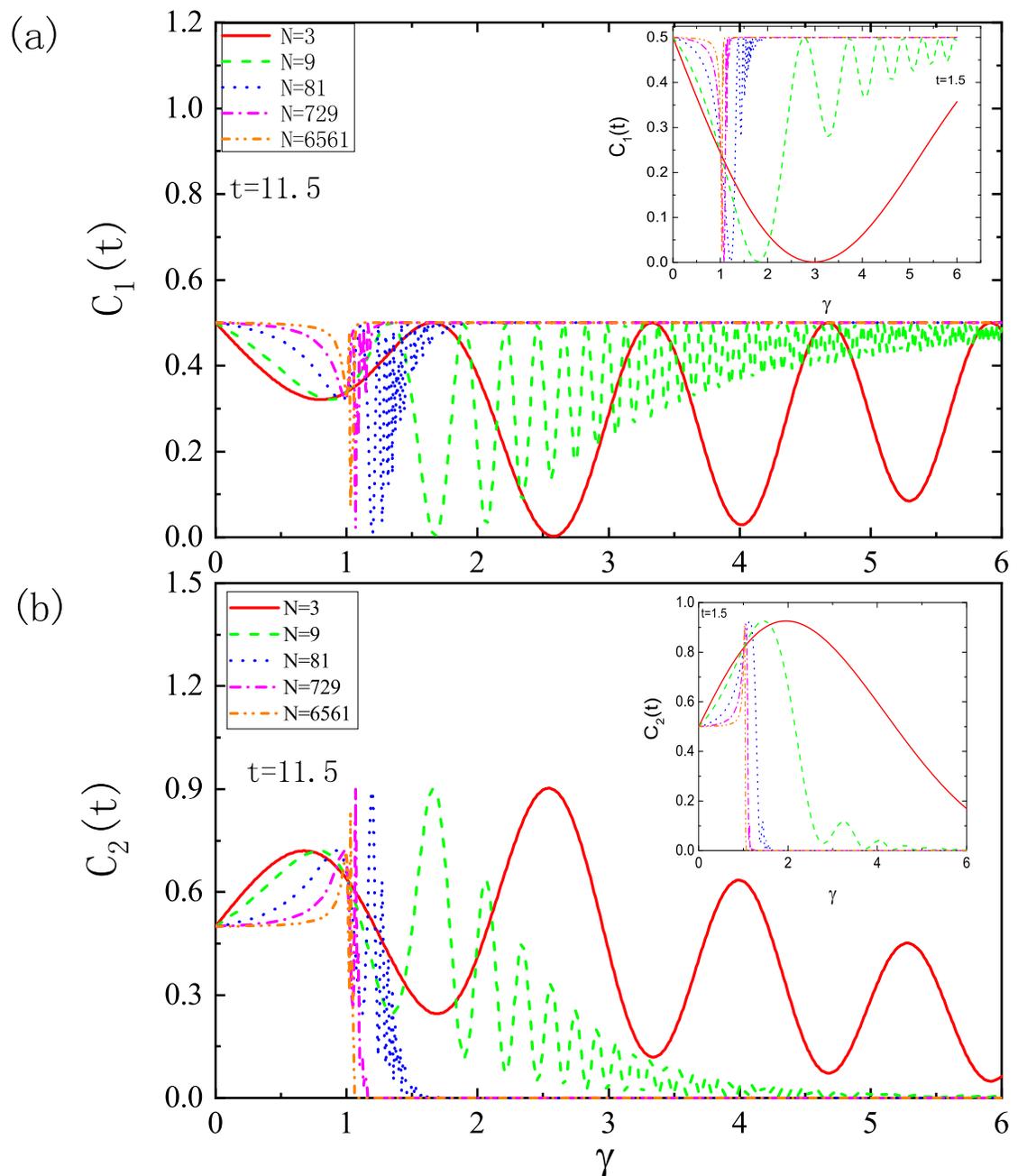
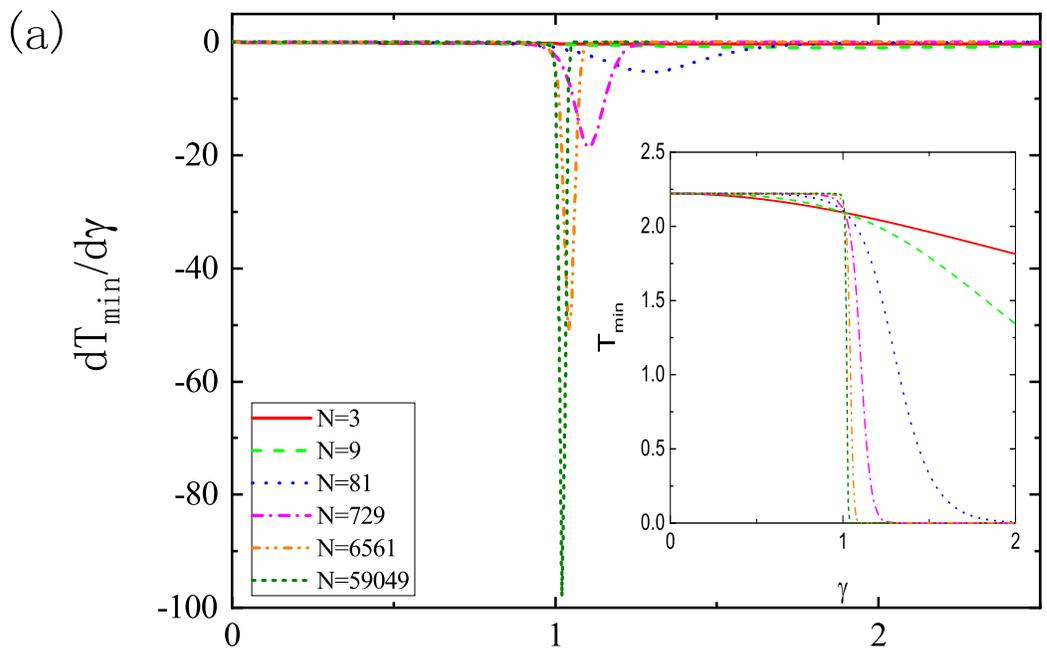
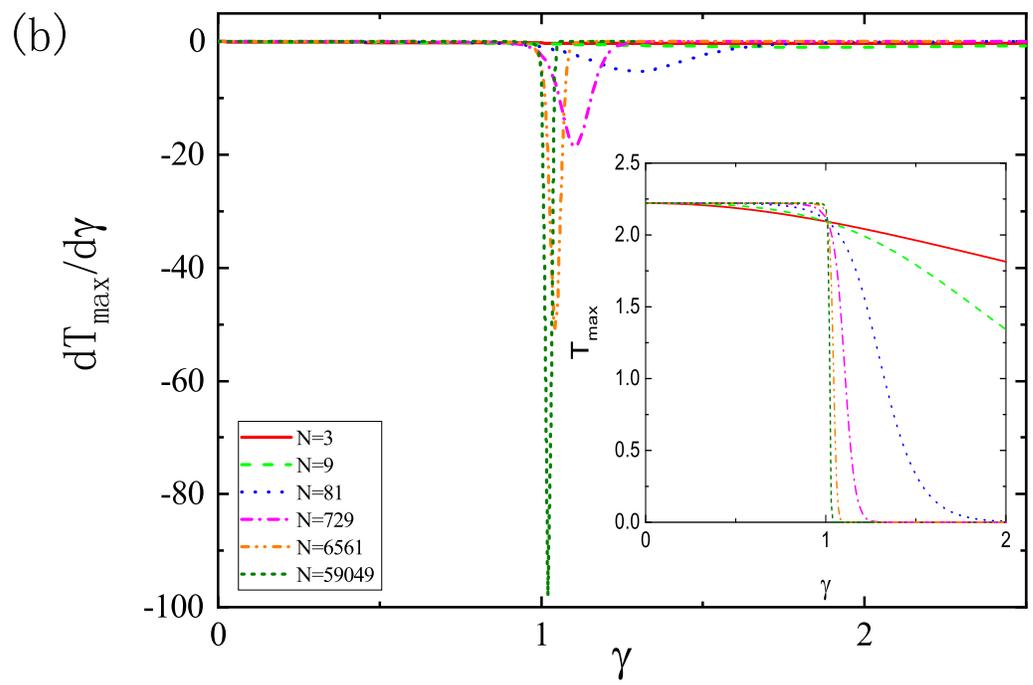
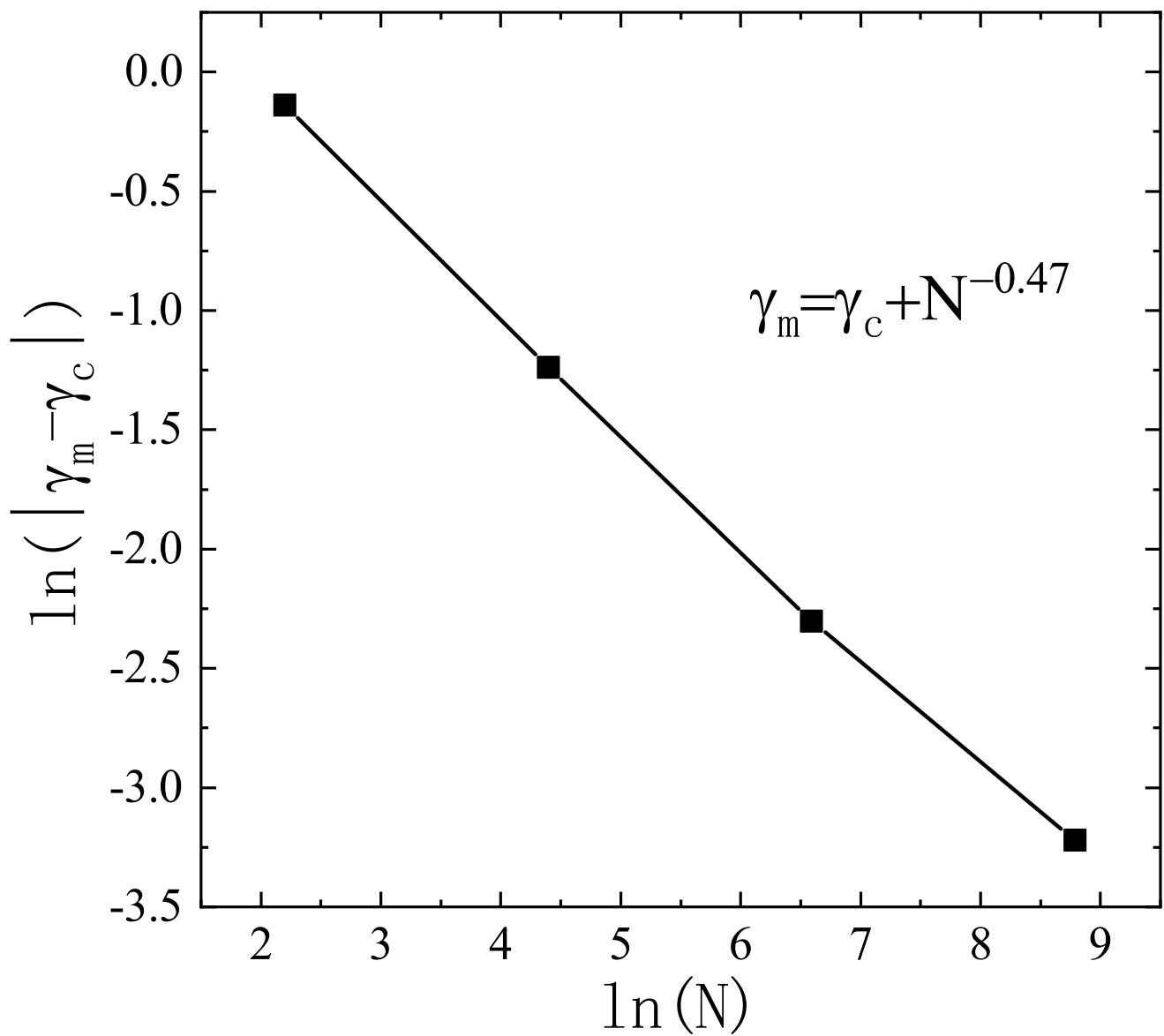
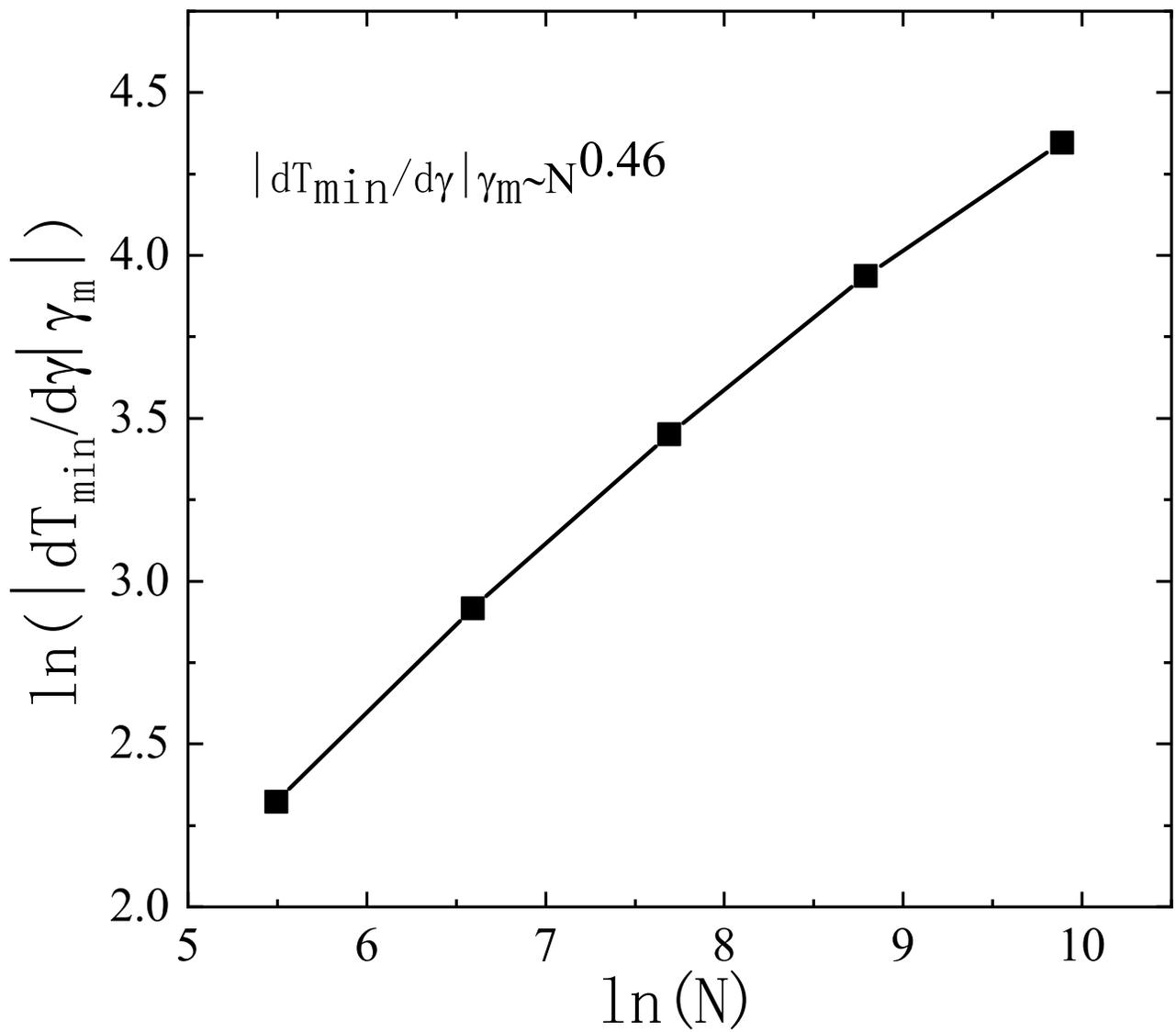
\end{document}